\documentclass[twocolumn,showpacs,preprintnumbers,prb,fleqn]{revtex4}
\usepackage{epsfig}
\usepackage{graphicx}
\usepackage{amsfonts}
\newcommand{\wt}{\widetilde}
\newcommand{\ig}{\includegraphics}
\begin{document}
                                                                                

\title{Critical behaviour of random fibers with mixed Weibull distribution}
\author{Uma Divakaran}
\email{udiva@iitk.ac.in}
\author{Amit Dutta}
\email{dutta@iitk.ac.in}
\affiliation{Department of Physics, Indian Institute of Technology Kanpur - 208016, India}
\date{\today}
\begin{abstract}
A random fiber bundle model with a mixed Weibull distribution 
is studied under the Global Load Sharing (GLS) scheme.  The mixed model
consists of two sets of fibers. The threshold strength of one set of fibers 
are randomly chosen from a Weibull 
distribution with a particular Weibull index, and another set of fibers with
a different index. The mixing tunes the critical stress of the bundle 
 and the variation of critical stress with the amount of
mixing is determined using a probabilistic method where the external
load is increased quasistatically. In a special 
case which we illustrate, the critical stress  is found to vary linearly 
with the mixing parameter. The critical exponents and power law behaviour
of burst avalanche size distribution is found to remain unaltered due to mixing.

\end{abstract}
\maketitle

\section{Introduction}
Sudden catastrophic failure of structures due to unexpected fracture of 
component materials is a concern and a challenging problem of physics as well
as engineering.
The dynamics of the failure of materials show interesting properties and hence 
there has
been an enormous amount of study on the breakdown phenomena till now\cite{Benguigui,Rava,zapperi}. 
The complexity involved in fracture processes can be suitably modelled
by grossly simplified models. The simplest
available model is Fiber Bundle Model(FBM)
\cite{Peirce,daniel,reviewchak, anderson}. 

A FBM consists of N parallel fibers.  
The disorder of a real system is introduced in the fiber bundle in the 
form of random distribution of strength of each fiber taken from a probability 
density p($\sigma$) and hence called Random Fiber Bundle Model (RFBM). 
The strength of each fiber is called its threshold 
strength. 
As a force F is applied externally on a bundle of N fibers, a stress 
$\sigma=F/N$
develops on each of them. The fibers which have their threshold strength
smaller than the stress generated, will break immediately. The next question 
that arises is the affect of breaking of these fibers on the remaining intact 
fibers $i.e.$, one has to decide a load sharing rule. 
The two extreme cases of load
sharing mechanisms are Global Load Sharing (GLS)\cite{daniel,dynamic,mixed3} 
and Local Load Sharing (LLS)\cite{Pacheco, Zhang,wu}.
In GLS, the stress of the
broken fiber is equally distributed to the remaining intact fibers. This rule
neglects local fluctuations in stress and therefore is effectively a 
mean field model
with long range interactions among the elements of the system \cite{Stanley}.
On the other hand, in LLS the stress of the broken fiber is given only 
to its nearest surviving neighbours. It is obvious that the actual breaking process
involves a sharing rule which is in between GLS and LLS. Several studies
have been made which considers a rule interpolating between GLS and LLS
\cite{mixed1,mixed2}.

For a given force F, some fibers break and they distribute their load to the
surviving fibers
following a load sharing rule causing further failure and redistribution
 of stress. This process continues until all the remaining fibers have 
their threshold strength greater than the redistributed stress
acting on them. This corresponds to the fixed point of the 
dynamics of the system.
As the applied force is increased on the system, more and more fibers break.
An avalanche of size S is defined as the number of failed fibers
between two successive external loadings.
There 
exist a critical load (or stress $\sigma_{c}$) beyond which if the load is 
applied, complete failure of the system takes place. Most of the studies on
FBM involve the determination of the critical stress $\sigma_{c}$ and the 
investigation of the type 
of phase transition from a state of partial failure to a state of complete 
failure. It has been shown that a bundle following GLS has a finite value 
of critical stress and belongs to a universality class with a 
specific set of critical exponents\cite{dynamic,pratip} 
whereas there is no finite critical stress $\sigma_{c}$ at thermodynamic 
limit in the case of LLS in one dimension\cite{Pacheco,Smith}.
On the other hand, LLS on complex network has been shown to belong to the 
same universality
 class as that of GLS with the same critical exponents\cite{complex}.

In this paper, we study a FBM with mixed fibers. Fibers have their threshold
strength randomly chosen from Weibull distributions with two different index
parameters. The motivation here is to study the dynamics of random fibers
in the presence of disorder caused due to mixing of two types of fibers with 
overlapping distribution of threshold strengths.
Moreover, the probabilistic method implemented to estimate the critical stress
can be very easily used to study any type of mixed fiber bundles
thus enabling one to put maximum disorder in the system being studied.

Section II consists of description of the model highlighting the method used.
In section III we present the results. Section IV includes discussions
and conclusions.

\section{The Model}

In the Weibull distribution (WD) of threshold strength of fibers, the 
probability of failure of each element when a stress $\sigma$ is generated
has a form
\begin{equation}
P(\sigma)=1-e^{-(\frac{\sigma}{\sigma_{0}})^{\rho}}
\end{equation}
where $\sigma_{0}$ is a reference strength and $\rho$ is called the Weibull 
index. We consider a mixed RFBM  with WD
of threshold strength of fibers where strengths of fibers are randomly
chosen from two different distributions characterised by different Weibull
indices and study the critical behaviour of the model. 
A fraction $x$ (henceforth called the mixing parameter) of fibers belong
to the class A (WD with $\rho=\rho_1$) and the remaining $(1-x)$ fraction belong
to the class B (WD with $\rho =\rho_2$)  
with the reference strength $\sigma_{0}$ set  equal to one
 in both the distributions.

The probabilistic method introduced by Moreno, Gomez and Pacheco
\cite{mixed3} 
is extended to explore the critical behaviour of the above model.
For the conventional WD $(x=1~\rm {or}~0)$, let us 
consider  a situation where the stress on each fiber increases from 
$\sigma_{1}$ to $\sigma_{2}$.  The probability that a fiber randomly chosen 
from the WD survives from the load $\sigma_{1}$ but fails 
when the load is $\sigma_{2}$, is given by
$$p(\sigma_{1},\sigma_{2})=\frac{P(\sigma_{2})-P(\sigma_{1})}{1-P(\sigma_{1})}=
1-e^{-(\sigma_{2}^{\rho}-\sigma_{1}^{\rho})}.$$
Thus the probability that the chosen fiber that has survived the load $\sigma_{1}$ also survives the (higher) load $\sigma_{2}$ is 
$q(\sigma_{1},\sigma_{2})=e^{-(\sigma_{2}^{\rho}-\sigma_{1}^{\rho})}$.
The key point is that the force F on the bundle is increased quasistatically
so that only the weakest fiber amongst the remaining intact fibers break.
Thus, one needs to identify the weakest fiber, break that fiber and then 
calculate the number of remaining unbroken fibers using $q(\sigma_1,\sigma_2)$
recursively, updating the value of $\sigma_2$ and $\sigma_1$ due to the 
load given away by the broken fibers, until no more failure occurs.
As mentioned previously, the dynamics of breaking will continue 
till the system reaches a fixed point. The process of slow increase of external
load is carried on up to the critical stress $\sigma_{c}$. The method avoids
the random averaging involved in a Monte Carlo simulation and hence turns out
to be very useful in dealing with fluctuations.

However, to deal with the
$'\rm mixed'$ random fiber bundle model, the  above method 
needs to be generalised in an appropriate manner.
The essential point is the fact that we need  to keep track of 
number of unbroken fibers in each distribution separately.
Let us assume that after a loading is done and a fixed point is reached, 
$N_{k_{1}}$ and $N_{k_{2}}$ are the number of unbroken fibers corresponding to 
$\rho=\rho_1$ and $\rho=\rho_2$ distribution respectively, 
and $\sigma_{k}$ is the stress per fiber at that instant.  
\begin{figure}[h]
\includegraphics[height=2.0in,width=2.1in]{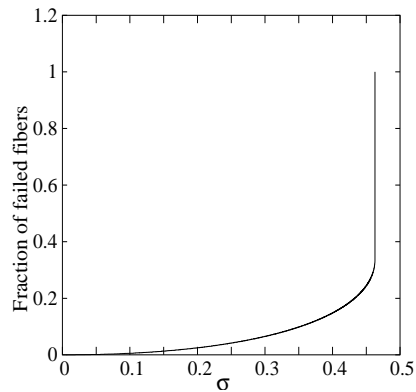}
\caption{Variation of fraction of broken fibers with external load
per fiber  $\sigma$ for a RFBM with two distributions corresponding 
to $\rho_1=2,~\rho_2=3$, $x=0.5$ and N=50000.}
\end{figure}

We define $q_{1}$ and $q_{2}$ as follows:
\begin{eqnarray}
q_{1}(\sigma_{1},\sigma_{2}) = e^{-(\sigma_{2}^{\rho_1}-\sigma_{1}^{\rho_1})}\nonumber\\
q_{2}(\sigma_{1},\sigma_{2}) = e^{-(\sigma_{2}^{\rho_2}-\sigma_{1}^{\rho_2})}\nonumber.
\end{eqnarray}
One needs to calculate the load
$(N_{k_{1}}+N_{k_{2}})\sigma_{l}$ that has to be applied to break
one fiber. Here, $\sigma_{l}={\rm min}(\sigma_{l_{1}},\sigma_{l_{2}})$
where $\sigma_{l_{1}}$ (or $\sigma_{l_{2}}$) is the next weakest fiber
in the $\rho=\rho_1$ (or $\rho=\rho_2$) distribution of strength of fibers 
and is obtained by the solution of the following equations:
\begin{eqnarray}
N_{k_{1}}-1 = N_{k_{1}}q_{1}(\sigma_{k},\sigma_{l_{1}})\nonumber\\
\noindent {\rm and}~~N_{k_{2}}-1 = N_{k_{2}}q_{2}(\sigma_{k},\sigma_{l_{2}})\nonumber
\end{eqnarray}
which gives
\begin{eqnarray}
\sigma_{l_{1}} = [\sigma_{k}^{\rho_1}-{\rm ln}(1-\frac{1}{N_{k_{1}}})]^{1/\rho_1}\\
\sigma_{l_{2}} = [\sigma_{k}^{\rho_2}-{\rm ln}(1-\frac{1}{N_{k_{2}}})]^{1/\rho_2}
\end{eqnarray}
The breaking of one fiber and the redistribution of its stress to all the
remaining intact fibers causes some
more failures. Let us assume that during this avalanche, at some point before 
the fixed point is reached, there are $\wt{N_{k_{1}}}$ and $\wt{N_{k_{2}}}$ 
number of unbroken fibers belonging to the two distributions where 
each fiber is under a stress $\wt{\sigma_{k}}$. This stress causes 
some more failures and as a result $N_{k_{1}}^{'}$ and $N_{k_{2}}^{'}$ fibers 
are unbroken. Let the new stress developed be $\sigma_{k}^{'}$.
The number of fibers which survive $\wt{\sigma_{k}}$ and $\sigma_{k}^{'}$
are obtained using the relation
\begin{eqnarray}
N_{k_{1}}^{''} = N_{k_{1}}^{'}q_{1}(\wt{\sigma_{k}},\sigma_{k}^{'})\\
N_{k_{2}}^{''} = N_{k_{2}}^{'}q_{2}(\wt{\sigma_{k}},\sigma_{k}^{'})
\end{eqnarray}
The stress on each fiber is now equal to 
$(N_{k_{2}}^{'}+N_{k_{1}}^{'})\sigma_{k}^{'}/(N_{k_{1}}^{''}+N_{k_{2}}^{''})$.
Eq (4) and Eq. (5) are used again and again until a fixed point is reached.
The fixed point condition is given by 
$(N_{k_{1}}^{'}+N_{k_{2}}^{'})-(N_{k_{1}}^{''}+N_{k_{2}}^{''})<\epsilon$ 
where $\epsilon$ is a small 
number (0.001). It should be noted that  the critical behaviour does 
not depend on 
the choice of $\epsilon$.
After the fixed point is reached, stress $\sigma_{l}$ is calculated once again as mentioned before and the whole process is repeated till complete failure
 occurs at $\sigma_{c}$.  

It should also be mentioned here that the critical stress $\sigma_c$ can also
be derived using directly  the cumulative distribution function for the  
mixed model as
given in Eq.~(6) in $p(\sigma_1,\sigma_2)$. However
in this case, the expression of $\sigma_l$, as defined above, turns out to be
very complicated and it is difficult to arrive at a simple closed form
as shown in Eq. (2) or (3).

\section{Results}
We present here the main results of a particular case, $x=0.5,~\rho_1=2$ and
$\rho_2=3$.
Fig.~1 shows the fraction of total number of broken fibers as 
a function of applied stress $\sigma$ . 
The graph clearly shows the existence
of a critical stress $\sigma_c = 0.46$ at which fraction of
failed fibers increases rapidly and the bundle breaks down completely.
The critical stress of the mixed bundle lies between that of the two 
pure bundles (for $\rho=2, ~\sigma_c=0.42$ and for $\rho=3,~\sigma_c=0.49$ 
obtained  using $\sigma_c=(\rho e)^{-1/\rho}$)\cite{daniel}.
Thus, the resulting 
critical stress of the mixed fiber bundle model can be tuned by varying the
mixing parameter $x$.

The mean avalanche size S of failure  is defined as the total number of 
broken fibers between
two successive loadings.
It diverges near the critical point as $(\sigma_c-\sigma)^{-\gamma}$
with an exponent $\gamma=1/2$. Scaling behaviour of S is shown in Fig.~2. 
\begin{figure}[h]
\includegraphics[height=2.1in,width=2.1in]{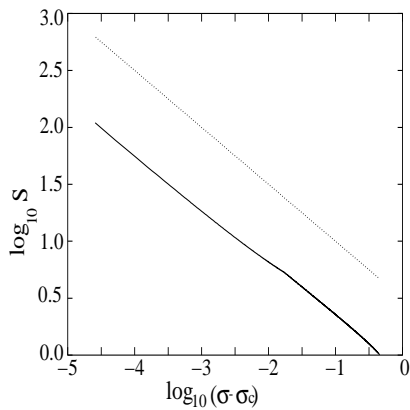}
\caption{Scaling behaviour of mean avalanche size S as the critical point
is reached for the mixed model. Also shown is a straight line (dotted) 
with slope (-1/2). Here, N=50000 and $x=0.5$.}
\end{figure}
\begin{figure}[h]
\ig[height=2.1in,width=2.1in]{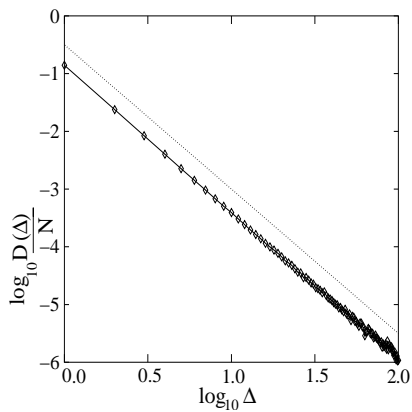}
\caption{Avalanche size distribution for x=0.5, $\rho_1=2$, $\rho_2=3$ and 
N=50000. A straight line with slope -5/2 is also shown.}
\end{figure}

The important feature associated with the failure process in RFBM is the
power law behaviour of burst avalanche distribution of fibers. 
The probabilistic approach \cite{mixed3} implemented to determine $\sigma_c$
turns out to be inappropriate in exploring the behaviour of avalanche size
distribution. This is due to the fact that the mean avalanche sizes ($S$) 
obtained 
for different $\sigma$
are of fractional values which leads to a difficulty in calculating
the distribution of avalanche sizes. We therefore use the standard Monte-Carlo
method along with the weakest fiber approach \cite{hansen} 
where the external load is
increased by an amount sufficient to break the weakest intact fiber.
The corresponding integer value of the size of an avalanche is denoted by
$\Delta$.
 For GLS, 
the distribution $D(\Delta)$ of an avalanche of size $\Delta$  
follows a power law
$D(\Delta)\propto\Delta^{-\xi}$, where $\xi=5/2$ in the asymptotic limit
\cite{hansen}.
Fig.~3 shows the avalanche size distribution for the present 
mixed RFBM obtained numerically with $x=0.5$; 
a power law behaviour with the same exponent
 5/2 is clearly observed,  confirming the mean field nature of the model.
That the 5/2 behaviour is expected even for a mixed RFBM for any $x$
can be justified using the saddle point method applicable in the limit of large 
$\Delta$\cite{hansen}.
In the present case ($\rho_1=2$ and $\rho_2=3$), the probability that a 
fiber will break when subjected to a stress $\sigma$ is
\begin{eqnarray}
P(\sigma)=x[1-{\rm exp}(-\sigma^2)]+(1-x)[1-{\rm exp}(-\sigma^3)]
\end{eqnarray}
so that the density distribution becomes
\begin{eqnarray}
p(\sigma)=2x\sigma {\rm exp}(-\sigma^2)+(1-x)3\sigma^2 {\rm exp}(-\sigma^3).
\end{eqnarray}
The avalanche size distribution takes the form
\begin{eqnarray}
\frac{D(\Delta)}{N}=\frac{\Delta^{\Delta-1}}{\Delta!}\int_{0}^
{\sigma^*}d\sigma\frac{1}{\sigma}[1-P(\sigma)-\sigma p(\sigma)]\nonumber\\
\times\left[\frac{\sigma p(\sigma)}{1-P(\sigma)}
{\rm exp}(-\frac{\sigma p(\sigma)}{1-P(\sigma)})\right]^{\Delta}
\end{eqnarray}
where $\sigma^*$ is the redistributed stress at the critical point
at which the average applied force $[F=N\sigma(1-P(\sigma))]$ maximises,
 and $p(\sigma)$ and
$P(\sigma)$ are as defined above.
The function inside the square bracket has a maximum when 
\begin{eqnarray}
\frac{\sigma p(\sigma)}{1-P(\sigma)}=1.
\end{eqnarray}
It should be noted that the above condition is satisfied when $\sigma=\sigma^*$.
Since the threshold distribution of fibers, (Eq.~(7)), does not have any
discontinuity, the saddle point integration of Eq.~(8) 
(retaining the first order term in the expansion of the prefactor
$(1-P(\sigma)-\sigma p(\sigma))$ around $\sigma=\sigma^*$) yields the 
asymptotic behaviour $D(\Delta)\propto \Delta^{-5/2}$.
\begin{figure}[h]
\includegraphics[height=2.1in,width=2.1in]{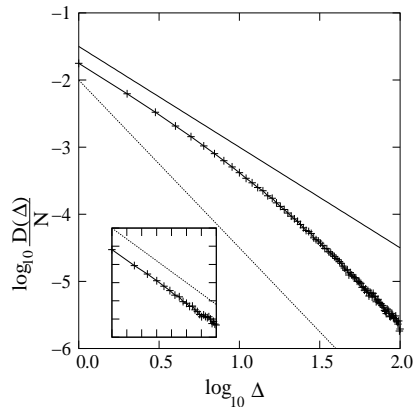}
\caption{ The variation of $D(\Delta)$ with $\Delta$ when close to the critical
distribution for a mixed Weibull distribution with 
$\rho_1 =2$, $\rho_2 =3$ and $x=0.5$. A clear crossover from 
$\Delta^{-3/2}$ to $\Delta^{-5/2}$ is observed with $\sigma_0=0.62$. The slope
of the thick line is -3/2 and dotted line is -5/2.
The inset shows the avalanche size distribution for the critical case along with
a line of slope of -3/2.}
\end{figure}

Let us now comment on the imminent failure \cite{pradhan05} behaviour, 
when a fraction 
of weak fibers are already removed, and the distribution is close to the 
critical distribution $i.~e.$, the strength of the weakest intact fiber
$(\sigma_0)$ is close to
the redistributed stress $\sigma^{*}$ at the critical point. 
We shall consider the case when $\rho_1 =2$ and $\rho_2 =3$. The variation
of $D(\Delta)$ with $\Delta$ is shown in Fig.~4 and as in the pure Weibull case,
we observe a crossover from $\Delta^{-3/2}$  to  $\Delta^{-5/2}$  as $\Delta$
increases. For the critical distribution, however, we observe a $\Delta^{-3/2}$
behaviour for the whole range of $\Delta$ (Inset Fig.~4). 

Although the power law behaviour of the avalanche size distribution near
the critical distribution remain unaltered by mixing, one may ask the
question about the behaviour of $\Delta_c$ with $x$
(where $\Delta_c$ denotes the
avalanche size at which crossover from 3/2 to 5/2 is observed). For the
case of interest, $\rho_1=2$ and $\rho_2=3$, $\Delta_c$ does not change
appreciably with $x$.  This is due to the fact 
that $\sigma^*$ is almost
constant as $x$ is varied (explained later). On the other hand, 
if we consider the case
$\rho_1=1$ and $\rho_2=3$ (Fig.~6), $\Delta_c$ should decrease with $x$
keeping $\sigma_0$ constant because as x is increased
$\sigma^*$ also increases taking the system away from critical distribution.

The critical exponent $\gamma$ and the power-law behaviour of avalanche size
distribution of
the mixed model remain unchanged for any value of mixing parameter $x$.
This supports the fact that the critical behaviour of a fiber bundle model is
determined entirely by the load-sharing rule.
\begin{figure}
\includegraphics[height=2.1in,width=2.1in]{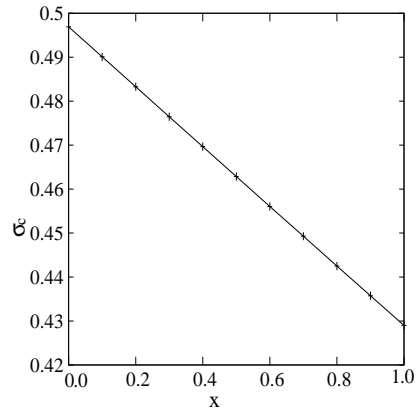}
\caption{Variation of critical stress $\sigma_{c}$ with $x$. Slope of
the straight line is 0.068. N=50000.}
\end{figure}
\begin{figure}[t]
\includegraphics[height=2.1in,width=2.1in]{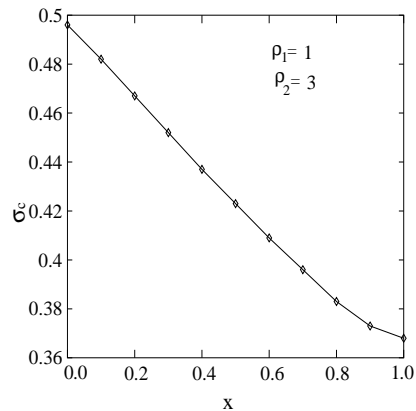}
\caption{Nonlinear variation of critical stress with $x$ for $\rho_1=1$ and
$\rho_2=3$.}
\end{figure}

However, the variation of $\sigma_c$ with the mixing parameter $x$ shows
a very interesting linear behaviour (Fig.~5).
To justify the linear behaviour, we recall the probability 
distribution (Eq.~(6)). 
A relation between the critical stress and $\sigma^*$ can be obtained
by calculating the applied force at $\sigma^*$ where $\sigma^*$ is a solution of
Eq.~9.
\begin{eqnarray}
\sigma_c=\frac{F}{N}=\sigma^{*^2} p(\sigma^*).
\end{eqnarray}

For the conventional WD,  
$\sigma^*=(\frac{1}{\rho})^\frac{1}{\rho}$, which immediately
reveals the fact that the difference in $\sigma^*$ for the cases $\rho=2$
and $\rho=3$ is very small as compared to the change in corresponding
$\sigma_c$ values. To a good approximation, one can set
$\sigma^*={\rm constant}=c$, so that Eq.(10) combined with Eq.~(7) yields
\begin{eqnarray}
\sigma_c&=&c^{2}[2xc{\rm exp}(-c^2)] + (1-x) 3c^2 {\rm exp}
(-c^3)\nonumber\\
&=&x c^2 [2c{\rm exp}(-c^2)-3c^2{\rm exp}(-c^3)] + 3c^4{\rm exp}(-c^3)\nonumber,
\end{eqnarray}
which justifies the dominant linear dependence of $\sigma_c$ on $x$
for this particular case. It should also be emphasised that this linear
relationship 
is a characteristic of situations where the variation of 
$\sigma^*$ is negligible
(to the lowest order) as $x$ is tuned from zero to one 
($e.g.,$ the present case)  and 
in general there is a non-linear relationship as shown in Fig.~6.

In a recent work \cite{raischel}, the shear failure of a glued interface
 has been
studied using a simple beam model where beams (fibers) connect the
two surfaces.  The stretching and bending threshold strength of the beams,
denoted by $\epsilon_1$ and $\epsilon_2$ respectively, 
are randomly distributed variables satisfying a joint probability distribution
function $p(\epsilon_1, \epsilon_2)$. 
The mean field critical exponents
are  obtained when the threshold distributions for bending and stretching
modes
are independent and chosen from two different WD.  In our model,
we study fibers with threshold strength chosen from two Weibull
distribution with different Weibull indices and 
 the critical exponents stick to
the mean field values also in our case.

\section{Conclusion}
The critical stress of a mixed RFBM with Weibull distribution and 
GLS is studied using a probabilistic 
approach where the external force is increased quasistatically at every
step of loading\cite{mixed3}. The advantage of this method is that it does not 
require the process of random averaging which takes comparatively 
longer computational time. 
We obtain the variation of critical stress with parameter $x$. This functional
dependence of the critical stress on the
characteristic quantity $'\rm {the~ mixing~ parameter}'$ is an important
objective of this work.

The critical behaviour of the mixed model namely
the critical exponents and the power law behaviour of the burst avalanche 
distribution are same as mean field. 
In a mixed Weibull distribution, the threshold
distribution of the two types of fibers are overlapping or the
distribution is continuous. The presence of discontinuity modifies the
avalanche size distribution for smaller $\Delta$\cite{udiva1}. Hence one expects
the mean field (GLS) behaviour of the mixed model. The behaviour of the
imminent failure shows a crossover in the avalanche size exponent from
5/2 to 3/2 as the critical distribution is approached.
In some distributions (depending upon $\rho_1$  and $\rho_2$), 
the effective critical stress
is found to vary linearly with the mixing parameter $x$. We have pointed out
the origin of the apparent linear behaviour and argued that in general a 
non-linear variation is expected.

\begin{center}
{\bf ACKNOWLEDGMENTS}
\end{center}
We sincerely thank Y. Moreno, A. F. Pacheco and S. Pradhan for their help and 
B. K. Chakrabarti for useful comments.


\begin{thebibliography}{05}

\bibitem{Benguigui}
H. J. Herrmann and S. Roux, {\it Statistical Models of Disordered Media}, North 
Holland, Amsterdam (1990);
B. K. Chakrabarti and L. G. Benguigui, {\it Statistical Physics of fracture and
Breakdown in Disordered Systems}, Oxford Univ. Press, Oxford (1997);
P. Bak, {\it How Nature Works}, Oxford Univ. Press, Oxford (1997);
M. Sahimi, {\it Heterogeneous Materials II: Nonlinear Breakdown Properties and 
Atomistic Modelling}, Springer-Verlag Heidelberg, (2003).

\bibitem{Rava}
R. da Silveria, Am. J. Phys. {\bf 67}, 1177 (1999).

\bibitem{zapperi}
S. Zapperi, P. Ray, H. E. Stanley and A. Vespignani, Phys. Rev. E {\bf 59}, 5049
(1999).

\bibitem{Peirce}
F. T. Peirce, J. Text. Inst. {\bf 17}, 355 (1926);
B. D. Coleman, J. Appl. Phys. {\bf 29}, 968 (1958).

\bibitem{daniel}
H. E. Daniels, Proc. R. Soc. London A {\bf 183}, 404 (1945).

\bibitem{anderson}
J. V. Andersen, D. Sornette and K. T. Leung, Phys. Rev. Lett. {\bf 78}, 2140 (1997)
.

\bibitem{reviewchak}
S. Pradhan and B. K. Chakrabarti, Int. J. Mod. Phys. B {\bf 17}, 5565 (2003);
P. C. Hemmer, A. Hansen, S. Pradhan, cond-mat/0602371.

\bibitem{dynamic}
S. Pradhan, P. Bhattacharyya and B.K. Chakrabarti, Phys. Rev. E {\bf 66},
016116 (2002).

\bibitem{mixed3}
Y. Moreno, J. B. Gomez and A. F. Pacheco, Phys. Rev. Lett. {\bf 85}, 2865 (2000).

\bibitem{Pacheco}
J. B. Gomez, D. Iniguez and A. F. Pacheco, Phys. Rev. Lett. {\bf 71}, 380 (1993).

\bibitem{Zhang}
S. D. Zhang, E-jiang  Ding, Phys. Rev. B {\bf 53}, 646 (1996).

\bibitem{wu}
B. Q. Wu and P. L. Leath, Phys. Rev. B {\bf 59}, 4002 (1999).

\bibitem{Stanley}
Compare Ising Model with infinite range interaction, see for example,
H. E. Stanley, {\it Introduction to Phase Transitions and Critical Phenomenon},
Oxford University Press, Oxford (1987).

\bibitem{mixed1}
R. C. Hidalgo, Y. Moreno, F. Kun and H.J. Herrmann, Phys. Rev. E {\bf 65}, 
046148 (2002).

\bibitem{mixed2}
S. Pradhan, B.K. Chakrabarti and A. Hansen, Phys. Rev. E {\bf 71}, 036149 (2005).

\bibitem{pratip}
P. Bhattacharyya, S. Pradhan and B.K. Chakrabarti, Phys. Rev. E {\bf 67},
046122 (2003).

\bibitem{hansen}
P. C. Hemmer and A. Hansen, J. Appl. Mech. {\bf 59}, 909 (1992);
A. Hansen and P. C. Hemmer, Phys. Lett. A {\bf 184}, 394 (1994).

\bibitem{Smith}
R. L. Smith, Proc. R. Soc. London A {\bf 372}, 539 (1980).

\bibitem{complex}
D. H. Kim, B. J. Kim and H. Jeong, Phys. Rev. Lett. {\bf 94}, 025501 (2005).

\bibitem{pradhan05} S. Pradhan, A. Hansen and P. C. Hemmer, Phys. Rev. Lett.
{\bf 95} 125501 (2005)

\bibitem{raischel}
F. Raischel, F. Kun and H. J. Herrmann, Phys. Rev. E {\bf 72}, 046126 (2005).

\bibitem{udiva1}
Uma Divakaran and Amit Dutta, cond-mat/0608223. 

\end{thebibliography}
\end{document}